\documentstyle[prb,aps,psfig]{revtex}
\begin{document}
\title{Acoustic measurement of the low-energy excitations in Nd$_{2-x}$Ce$_{x}$CuO$%
_{4+\delta }$}
\author{F.Cordero,$^{1}$ C. R. Grandini,$^{2}$ and M. Ferretti$^{3}$}
\address{$^{1}$CNR, Istituto di Acustica ``O.M. Corbino``, Via del Fosso del Cavaliere 100,\\
I-00133 Roma, Italy and Unit\`{a} INFM-Roma1, P.le A. Moro 2,I-00185 Roma,Italy}
\address{$^{2}$Universidade Estadual Paulista, Departamento de Fisica, 17.033-360,Bauru, SP, Brazil}
\address{$^{3}$Universit\`{a} di Genova, Dipartimento di Chimica e Chimica Industriale,\\
and Unit\`{a} INFM-Genova, Via Dodecanneso 31, I-16146 Genova, Italy}
\maketitle

\begin{abstract}
The complex dynamic Young's modulus of ceramic Nd$_{2-x}$Ce$_{x}$CuO$_{4}$
with $x=0,$ 0.05 and 0.20 has been measured from 1.5 to 100~K at frequencies
of $1-10$~kHz. In the undoped sample the modulus starts decreasing below $%
\sim 20$~K, instead of approaching a constant value as in a normal solid.
The modulus minimum has been interpreted in terms of paraelastic
contribution from the relaxation of the Nd$^{3+}$ $4f$ electrons between the
levels of the ground state doublet, which is split by the interaction with
the antiferromagnetically ordered Cu sublattice. The value of the splitting
is found to be 0.34~meV, in excellent agreement with inelastic neutron
scattering, infrared and specific heat experiments. With doping, the anomaly
shifts to lower temperature and decreases in amplitude, consistently with a
reduction of the local field from the Cu sublattice.
\end{abstract}

%\draft

\twocolumn

\section{Introduction}

The crystal-field (CF) excitations of the Nd$^{3+}$ $4f$ electrons in Nd$%
_{2-x}$Ce$_{x}$CuO$_{4}$ have been extensively studied, mainly by inelastic
neutron scattering (INS) \cite{HF98,SAF90,BDP92,HRF98} and infrared spectroscopies.
\cite{JRP00} Although discrepancies exist between the proposed sets of
phenomenological parameters in the CF hamiltonian, there is general
agreement on the level scheme.\cite{JRP00,HDC98,HF98} The ground state
multiplet of Nd$^{3+}$ has $J=9/2$ and is split into five doublets, the
separation of the first two being 14~meV; each doublet, in turn, is split by
the interaction of the Nd magnetic moment with the local magnetic field,
mainly created by the antiferromagnetic ordering of the Cu magnetic moments
below $T_{\text{N}}^{\text{Cu}}\simeq 260$~K. The magnetic separation of the
first doublet has been determined by INS, and passes from 0.39~meV for the
undoped case to 0.2~meV for $x=0.15$.\cite{HDC98} Such a low-energy
excitation also appears as a Schottky anomaly in the low-temperature
specific heat,\cite{HDC98,AAC92,CDA92,BSR93,MMS97} with values of the level
splitting in agreement with the INS results; in principle, it should also
give rise to an anomaly in the elastic moduli, if the level separation can
be modulated by strain. To our knowledge, only two experiments have been
reported on the elastic properties of NCCO at liquid He temperatures,\cite
{FZZ91,STP90} but no attempts have been made to relate those data to the
elastic response of the ground state doublet. We made anelastic spectroscopy
measurements in the kHz range on Nd$_{2-x}$Ce$_{x}$CuO$_{4}$ with $x=0,$
0.05 and 0.20, and found a decrease of the Young's modulus at liquid He
temperatures, which can be identified with the paraelastic response of the
ground state doublet of Nd$^{3+}$.

\section{Experimental}

The samples were prepared by solid state reaction, with repeated sintering
and grinding of the powders up to obtaining X-ray diffraction spectra free
of unwanted peaks; the methods used, together with a thorough thermodynamic
study of the phase diagram and stability of the Nd-Ce-Cu-O system, are
reported in Ref. \onlinecite{DFF96}. Generally, a reducing treatment is made
on Nd$_{2-x}$Ce$_{x}$CuO$_{4}$, in order to make it superconducting; the
data presented here refer to the as-prepared or reoxygenated status.

The samples were cut as bars approximately $40\times 4\times 0.6$~mm$^{3}$.
The dynamic Young's modulus, $E\left( \omega ,T\right) =E^{\prime
}+iE^{\prime \prime },$ was measured by suspending the bars on their nodal
lines on thin wires and electrostatically exciting their flexural
vibrations; the 1st and 5th flexural modes were chosen, whose resonance
frequencies $\omega /2\pi $ are approximately in the ratio $1:13.3$. The
vibration amplitude was measured by a frequency modulation technique, where
the capacity between the sample and the electrode is inserted into a high
frequency oscillator, and the resulting signal, after passing through an
amplitude-modulation discriminator, is sent to a lock-in amplifier. The real
part $E^{\prime }$ of the Young's modulus is given by $E=\rho \left( 0.975%
\frac{\omega }{2\pi }\frac{l^{2}}{h}\right) ^{2}$, where $\rho $, $l$ and $h$
are the sample density, length and thickness; a correction for the porosity
should be necessary, but we are only interested in the relative variation of 
$E$ with temperature. The elastic energy loss coefficient $Q^{-1}=E^{\prime
\prime }/E^{\prime }$ was measured from the free decay of the vibration
amplitude after switching off the excitation signal.

\section{Results}

Figure 1 presents the complex Young's modulus of Nd$_{2-x}$Ce$_{x}$CuO$_{4}$
at three doping levels, between 1 and 100~K; only the data of the first mode
are presented, since those at higher frequency are identical. The elastic
energy loss coefficient decreases monotonously with temperature and with
increasing doping, without any particular feature except for a small peak at
15~K for the undoped sample. Such a peak disappears after annealing in
vacuum at high temperature, and, although several reorientational
transitions of the Nd and Cu spins have been observed at low temperature
with various techniques,\cite{HDC98,MYK90,KHS97} it does not seem to be
associated to any of these magnetic transitions. We will not discuss this
peak further. The elastic modulus presents a nearly linear stiffening on
cooling below room temperature, which levels off below 30~K; contrary to a
normal solid, however, the modulus does not stabilize on further cooling,
but drops of several parts in $10^{3}$. The drop actually saturates at the
lowest temperatures reached in the experiments (see also Fig. 2), and in
some cases we measured the beginning of a rise below 1.5~K. The temperature
and amplitude of the drop decrease with increasing doping similarly to the
Schottky anomaly in the specific heat, and therefore this feature might well
be connected with the ground state doublet of the Nd$^{3+}$ ions.

\section{Discussion}

Most of the specific heat data below 10~K have been interpreted in terms of
a Schottky anomaly arising from the ground state doublet,\cite
{HDC98,AAC92,CDA92,BSR93,MMS97} with a splitting $\Delta \left( T\right) =$ $%
2\mu _{0}B_{\text{Cu}}\left( T\right) $ determined through Nd-Cu exchange by
the staggered magnetization $B_{\text{Cu}}\left( T\right) $ of the AF Cu
sublattice below $T_{\text{N}}^{\text{Cu}}$; since $T_{\text{N}}^{\text{Cu}%
}\simeq 260$~K, $B_{\text{Cu}} $ and the splitting $\Delta $
are practically constant below 50~K. The existence of two levels $E_{i}$
separated by $\Delta $, whose internal energy is $U=$ $\left( \Delta
/2\right) \tanh \left( \Delta /2k_{\text{B}}T\right) $, contributes to the
molar specific heat with the Schottky term 
\begin{equation}
\delta C_{p}=\frac{\partial U}{\partial T}=k_{\text{B}}c\left( \frac{\Delta 
}{2k_{\text{B}}T}\right) ^{2}\frac{1}{\cosh ^{2}\left( \Delta /2k_{\text{B}%
}T\right) }\,,  \label{Cp}
\end{equation}
where $c$ is the molar concentration of the relaxing entities, in the
present case the concentration $1-x/2$ of Nd ions. On the other hand, if the
splitting can be modulated by a strain $\varepsilon $, paraelastic and
diaelastic\cite{LB1,CCC98} contributions to the elastic modulus $M=\partial
^{2}U/\partial \varepsilon ^{2}$ are expected. Here we omit the tensorial
character of the elastic properties, since we are dealing with the Young's
modulus of polycrystalline samples, which contains a combination of several
elastic constants and we are not entering in the details of the effect of
the strain symmetry on the Nd$^{3+}$ multiplets; then $M$ is identified with
the Young's modulus $E$ and $\varepsilon $ is the corresponding effective
strain.

The paraelastic contribution can be written as\cite{LB1,CCC98} $\delta M_{%
\text{para}}=\sum_{ij}\frac{\partial n_{i}}{\partial E_{j}}\frac{\partial
E_{i}}{\partial \varepsilon }\frac{\partial E_{j}}{\partial \varepsilon }$
and is due to the relaxation of the populations $n_{i}$, perturbed by the
vibration strain $\varepsilon $ through the modulation of the energies $%
E_{i} $. For only two levels $E_{2}-E_{1}=\Delta $, one obtains\cite{CCC98} 
\begin{equation}
\delta M_{\text{para}}=-\frac{c}{v_{0}}\gamma ^{2}\frac{1}{4k_{\text{B}%
}T\cosh ^{2}\left( \Delta /2k_{\text{B}}T\right) }\,,  \label{Mp}
\end{equation}
where $v_{0}$ is the molecular volume and $\gamma =\partial \Delta /\partial
\varepsilon $ is the deformation potential. Such a relaxation occurs with a
characteristic rate $\tau ^{-1}$, and therefore the dynamic modulus\cite{NB}
has to be multiplied by a factor $\left( 1+i\omega \tau \right) ^{-1}$. The
resulting real part of the dynamic modulus $M^{\prime }\left( \omega
,T\right) $ acquires a frequency-dependent dispersion, while the imaginary
part produces absorption in correspondence with the modulus dispersion. The
present data, however, do not exhibit any frequency dependence of $M^{\prime
}\left( \omega ,T\right) $ or absorption peak, and this implies $\omega \tau
\simeq 0$, namely the rate $\tau ^{-1}$ is much faster than the measuring
frequencies $\omega \sim 10^{4}-10^{5}$~s$^{-1}$. Indeed, the Cu and Nd spin
rates deduced from muon spin relaxation experiments\cite{HMK97} in samples
with $x=0.20$ are faster than $10^{10}$~s$^{-1}$.

If the first derivative $\gamma $ of the energy split with respect to strain
is null for symmetry reasons, than one remains with the diaelastic term, $%
\delta M^{\text{dia}}=\frac{c}{v_{0}}\sum_{i}n_{i}\frac{\partial ^{2}E_{i}}{%
\partial \varepsilon ^{2}}$, which contains the second derivatives. The
physical meaning of such a term is simply that the curvature of each energy
level versus strain contributes to the corresponding elastic constant, and
its contribution is weighted with the level population. For only two levels
one obtains:\cite{CCC98} 
\begin{equation}
\delta M_{\text{dia}}=\frac{c}{v_{0}}\frac{1}{2}\frac{\partial ^{2}\Delta }{%
\partial \varepsilon ^{2}}\tanh \left( \frac{\Delta }{2k_{\text{B}}T}\right)
\,.  \label{Mdia}
\end{equation}

The continuous curves in Fig. 2 are fits with $\delta M_{\text{para}}$ given
by Eq. \ref{Mp} and assuming that the background modulus depends on
temperature like an insulating Debye solid,\cite{Ale66} $M\left(
T\right) =M_{0}\left( 1-aT^{4}\right) $. An additional term quadratic in $T$
is expected when the system becomes metallic, due to the energy of the free
electrons, which increases quadratically with $T$; such a term, however,
could be of importance only for the highest doping $x=0.20$, and still at $%
x=0.15$ some elastic constants have been found to be well described by the
quartic term alone below 60~K.\cite{STP90} From the data at $x=0$ and 0.05 we
find that $M\left( T\right) $ is described by the coefficient $a=-2.1\times
10^{-10}$~K$^{-4}$. The values of the splitting deduced from the fits of
Fig. 2 are $\Delta /k_{\text{B}}=4.0$, 2.25 and 1.3~K at $x=0,$ 0.05 and
0.20 respectively, while the amplitudes of the paraelastic contributions are
in the ratios $1:0.6:0.05$. The splitting found for the undoped case is in
excellent agreement with the values deduced from INS\cite{SAF90} ($\Delta
/k_{\text{B}}=4.06$~K ), infrared spectroscopy\cite{JRP00} (3.8~K) and
specific heat\cite{HDC98} ($\Delta /k_{\text{B}}=4.5$~K). Doping by Ce
substitution introduces electrons mainly of Cu $3d$ character, which cancel
the magnetic moments of those Cu ions and therefore reduce the Cu-Nd
exchange responsible for the ground state splitting.\cite{BSR93} As a
result, the Schottky anomaly in the specific heat has been found to shift to
lower temperature, to broaden and to decrease in intensity with doping.\cite
{HDC98,BSR93,HRF98} The present data confirm such a trend, but the reduction
in intensity is more marked than in the specific heat measurements;\cite
{HDC98} possibly because of a reduction of the strain dependence of the
splitting with doping.

At $x>0.15$ there is no long range AF order of the Cu spins, and the
magnetic ground state splitting should disappear; nonetheless, the specific
heat presents a peak up to $x=0.20$.\cite{MMS97} Our data at $x=0.20$ also
present a residue due to the doublet with reduced splitting, but the value
deduced from the fit, $\Delta /k_{\text{B}}=1.2$~K, should be considered
with caution in view of the possible influence of an electronic quadratic
term in the temperature dependence of the modulus.

In some cases the fits to the specific heat data have been improved by
assuming an additional temperature dependence of the magnetic splitting
attributed to the ordering of the Nd momenta,\cite{HDC98,AAC92} which
however affects the splitting by less than 10\%, or by assuming a
three-level structure instead of a doublet.\cite{CDA92} We chose to not
overinterpret our data with additional parameters besides a simple Schottky
anomaly with constant splitting below 50~K and the $T^{4}$ dependence of the
background elastic modulus. In fact, the contribution of the doublet to the
elastic response, Eq. \ref{Mp}, has a slower temperature dependence than the
contribution to the specific heat, Eq. (\ref{Cp}), in the high temperature limit,
being proportional to $T^{-1}$ instead of $T^{-2}$; therefore, the elastic
anomaly is spread to higher temperature where additional effects become
important. In addition, the lowest temperature attained in our experiments
was not sufficient to complete the low-temperature side of the curves, so
that some uncertainty in the determination of the minimum of $\delta M_{%
\text{para}}\left( T\right) $ exists. This means that, while it is easy to
improve the fit by adding a temperature dependence of the splitting, by
taking into account the contribution of the excited doublets, and adopting
more sophisticated expressions for $M\left( T\right) $, it is difficult to
distinguish which of the improvements is the important one. Still, the data
presented here, together with other results on reduced samples, are
sufficient to indicate the presence of a minimum in the modulus due to a
paraelastic response, rather than a step-like anomaly due to a diaelastic
response. Fits to the dialeastic expression, Eq. \ref{Mdia}, are less
satisfactory and yield values of $\Delta $ about twice larger than those of
Fig. 2, in contrast with the literature values. If both the diaelastic and
paraelastic terms are considered, with both $\gamma $ and $\partial
^{2}\Delta /\partial \varepsilon ^{2}$ as free parameters, the splitting $%
\Delta $ remains almost the same as reported here and up to half of the anomaly
can be accounted for by the diaelastic term, without a significant improvement
of the fit. We conclude that the anelastic response in the modulus is mainly
due to the linear part of the response of the ground-state splitting to strain.
The magnitude of the deformation potential $\gamma \ $can be extracted
inserting in Eq. \ref{Mp} the value obtained from the fit, $\left( \delta M_{%
\text{para}}/M\right) _{\text{max}}=0.0044$ at 2~K for $x=0$, and using $%
M=100$~GPa, as determined from the resonance frequency (without corrections
for the porosity); it turns out $\gamma =0.013$~eV for $x=0$ and 0.01~eV for 
$x=0.05$.

As noted above, the absence of a rise of the absorption in correspondence to
the decrease of the modulus implies that the relaxation rate for the Nd$%
^{3+} $ spins to redistribute themselves within the ground states doublet is
much faster than our measuring frequency, $\omega \sim 10^{4}-10^{5}$~s$^{-1}$%
. Also, there are no signs of freezing of the Cu spins into a spin-glass or
cluster glass states, which instead is observed in the hole-doped CuO$_{2}$
planes of La$_{2-x}$Sr$_{x}$CuO$_{4}$ and YBa$_{2}$Cu$_{3}$O$_{6+x}$;\cite
{Joh97} in La$_{2-x}$Sr$_{x}$CuO$_{4}$ the frozen spin domains produce a
distinct rise of the acoustic absorption,\cite{79} which is totally absent
here. Similar conclusions have been drawn from $\mu $SR relaxation
experiments.\cite{HMK97}

To our knowledge, only two other experiments\cite{STP90,FZZ91} exist on the
acoustic properties of Nd$_{2-x}$Ce$_{x}$CuO$_{4}$ at liquid He
temperatures, which however cannot be compared with the present one in a
straightforward way; in fact, they report some elastic constants of single
crystals, while we measured the Young's modulus of polycrystals, which
contain a combination of all the compliances.\cite{NB} Fill {\it et al.}\cite
{FZZ91} found several anomalies in the elastic constants of undoped Nd$_{2}$%
CuO$_{4}$ as a function of temperature and magnetic field. In particular,
the $c_{66}$ shear presented a minimum at 5~K with an amplitude nearly ten
times larger than our minimum, and with a narrower shape, followed by a
smaller decrease below 1~K tentatively attributed to a Schottky anomaly or
magnetic ordering of Nd ions. Around 5~K also step-like changes have been
found in other elastic constants, which have been interpreted as a
ferroelastic transition driven by an ordering of the Nd spins.\cite
{FZZ91,KHS97} We cannot find a clear correspondence between those data and
the ones presented here.

Regarding the doped material, Saint-Paul {\it et al.}\cite{STP90} measured
the sound velocity of a single crystal of Nd$_{1.85}$Ce$_{0.15}$CuO$_{4}$;
the shear mode in the $ab$ plane did not present any anomaly down to 10~K,
but this is not in contrast with our data at $x=0.20$, where the decrease of
the modulus is hardly detectable above 10~K. Instead, the $c_{33}$ mode
exhibited an upward deviation below 10~K with respect to the $-T^{4}$
dependence, opposite to our measurements. A\ possible explanation for the
discrepancy is that the major contribution to the $c_{33}$ mode is
diaelastic with positive $\partial ^{2}\Delta /\partial \varepsilon ^{2}$,
while the Young's modulus of the polycrystal contains a predominant
paraelastic contribution which cancels the diaelastic one.

\section{Conclusion}

The Young's modulus and elastic energy loss coefficient of Nd$_{2-x}$Ce$_{x}$%
CuO$_{4}$ with $x=0,$ 0.05 and 0.20 have been measured from 1.5 to 100~K. On
approaching the lowest temperature, the modulus presents a drop whose
temperature and amplitude decrease with increasing doping. The data have
been interpreted in terms of paraelastic contribution from the relaxation of
the Nd$^{3+}$ $4f$ electrons between the levels of the ground state doublet,
which is split by the interaction with the antiferromagnetically ordered Cu$%
^{2+}$ ions. Excellent agreement is found with the value of the splitting at
zero doping deduced from INS, infrared and specific heat measurements, and
the shift of the anomaly to lower temperature with doping is consistent with
a reduction of the local field from the Cu sublattice. An effective
deformation potential $\gamma =0.013$~eV is found for the strain derivative
of the splitting at $x=0$, which decreases with doping. The absorption is
very low and featureless, indicating that the relaxation rate for reaching
equilibrium within the ground state doublet is much faster than the
measuring frequency, $\tau ^{-1}\gg 10^{5}$~s$^{-1}$, and also excluding the
occurrence of freezing phenomena of the Cu or Nd spins down to 1.5~K.

\section{Acknowledgments}

This work has been carried out withthin the framework of the CNR-CNPq
cooperation project 2001-2002.

%Captions to figures 
%una riga di testo qui è necessaria per SW, che altrimenti non carica il documento

\begin{figure}[tbp]
\caption{Elastic energy loss coefficient and relative change of the Young's
modulus $E$ of Nd$_{2-x}$Ce$_{x}$CuO$_{4}$ samples with $x=0$ (0.9 kHz), $%
x=0.05$ (2 kHz) and $x=0.20$ (1.6 kHz).}
\label{Fig1}
\end{figure}

\begin{figure}[tbp]
\caption{Fit of the relative change of the Young's modulus of Nd$_{2-x}$Ce$%
_{x}$CuO$_{4}$ with the paraelastic contribution from a doublet with
splitting $\Delta$, Eq. \ref{Mp}.}
\label{Fig2}
\end{figure}


\begin{references}

\bibitem{HF98}
W. Henggeler and A. Furrer, J. Phys.: Condens. Matter {\bf 10}, 2579-2596 (1998).

\bibitem{SAF90}
U. Staub, P. Allenspach, A. Furrer, S.-W. Cheong and Z. Fisk, Solid State 
Commun. {\bf 75}, 431 (1990).

\bibitem{BDP92}
A.T. Boothroyd, S.M. Doyle, D.MK. Paul and R. Osborn, Phys. Rev. B {\bf 45}, 
10075 (1992).

\bibitem{HRF98}
W. Henggeler, B. Roessli, A. Furrer, P. Vorderwisch and T. Chatterji, Phys. 
Rev. Lett. {\bf 80}, 1300 (1998).

\bibitem{JRP00}
S. Jandl, P. Richard, M. Poirier, V. Nekvasil, A.A. Nugroho, A.A. Menovsky, 
D.I. Zhigunov, S.N. Barilo and S.V. Shiryaev, Phys. Rev. B {\bf 61}, 12882 
(2000).

\bibitem{HDC98}
N.T. Hien, V.H.M. Duijn, J.H.P. Colpa, J.J.M. Franse and A.A. Menovsky, Phys. 
Rev. B {\bf 57}, 5906 (1998).

\bibitem{AAC92}
P. Adelmann, R. Ahrens, G. Czjzek, G. Roth, H. Schmidt and C. Steinleitner, 
Phys. Rev. B {\bf 46}, 3619 (1992).

\bibitem{CDA92}
S.J. Collocott, R. Driver and C. Andrikidis, Phys. Rev. B {\bf 45}, 945 (1992).

\bibitem{BSR93}
T. Brugger, T. Schreiner, G. Roth, P. Adelmann and G. Czjzek, Phys. Rev. Lett. 
{\bf 71}, 2481 (1993).

\bibitem{MMS97}
E. Maiser, W. Mexner, R. Sch{\"a}fer, T. Schreiner, P. Adelmann, G. Czjzek, 
J.L. Peng and R.L. Greene, Phys. Rev. B {\bf 56}, 12961 (1997).

\bibitem{FZZ91}
V.D. Fil, G.A. Zvyagina, S.V. Zherlitsyn, I.M. Vitebsky, V.L. Sobolev, S.N. 
Barilo and D.I. Zhigunov, Mod. Phys. Lett. B {\bf 5}, 1367 (1991).

\bibitem{STP90}
M. Saint-Paul, J.L. Tholence, S. Pi{\~n}ol, X. Obradors, R.J. Melville and S.B. 
Palmer, Solid State Commun. {\bf 76}, 1257 (1990).

\bibitem{DFF96}
M. Daturi, M. Ferretti, E.A. Franceschi and M. Minguzzi, Physica C {\bf 268}, 
300 (1996).

\bibitem{MYK90}
M. Matsuda, K. Yamada, K. Kakurai, H. Kadowaki, T.R. Thurston, E. Endoh, Y. 
Hidaka, R.J. Birgeneau, M.A. Kastner, P.M. Gehring, A.H. Moudden and G. Shirane,
 Phys. Rev. B {\bf 42}, 10098 (1990).

\bibitem{KHS97}
A.N. Knigavko, H.L. Huang and V.L. Sobolev, J. Appl. Phys. {\bf 81}, 4154 (1997).

\bibitem{LB1}
G. Leibfried and N. Breuer, {\it Point Defects in Metals I}.  (Springer, , 1978).

\bibitem{CCC98}
G. Cannelli, R. Cantelli, F. Cordero and F. Trequattrini, {\it Tunneling 
Sistems in Amorphous and Crystalline Solids}. ed. by P. Esquinazi, p. 389 
(Springer, Berlin,  1998).

\bibitem{NB}
A.S. Nowick and B.S. Berry, {\it Anelastic Relaxation in Crystalline Solids}.  
(Academic Press, New York, 1972).

\bibitem{HMK97}
M. Hillberg, M.A.C. de Melo, H.H. Klau{\ss}, W. Wagener, F.J. Litterst, P. 
Adelmann and G. Czjzek, Hyperfine Interact. {\bf 104}, 221 (1997).

\bibitem{Ale66}
G.A. Alers, {\it Physical Acoustics Vol. {IV A}}. ed. by W.P. Mason, p. 277-297 
(Academic Press, New York, 1966).

\bibitem{Joh97}
D.C. Johnston, {\it Handbook of Magnetic Materials}. ed. by K.H.J. Buschow, p. 
1 (North Holland,  1997).

\bibitem{79}
F. Cordero, A. Paolone, R. Cantelli and M. Ferretti, Phys. Rev. B {\bf 62}, 
5309 (2000).

\end{references}
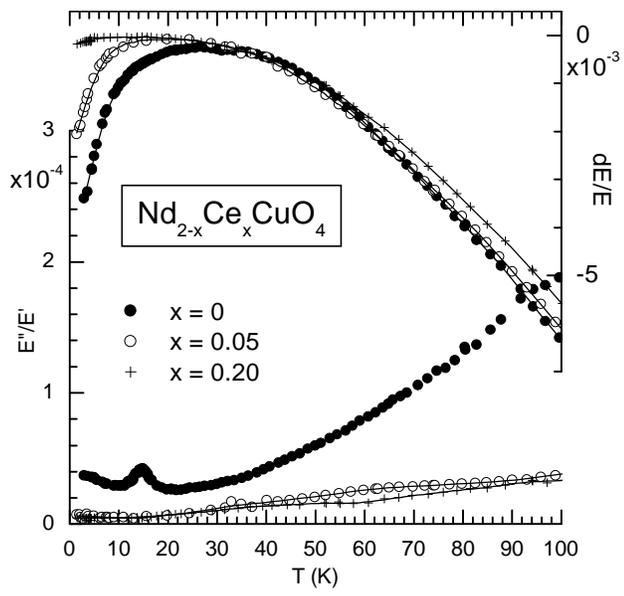
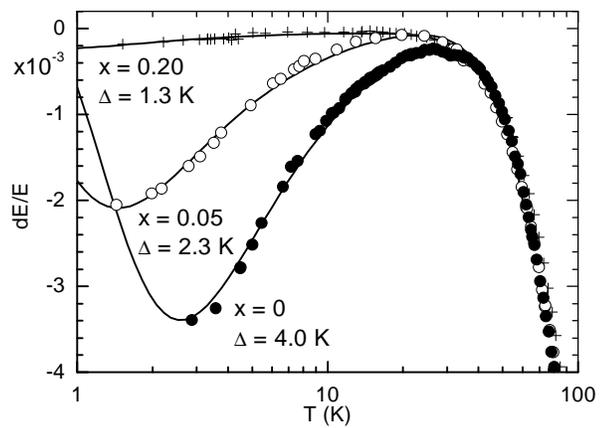
\end{document}